\documentclass[pss,fleqn]{w-art}
\usepackage{times}
\usepackage{w-thm}
\usepackage[]{graphicx}
\begin{document}
\DOIsuffix{theDOIsuffix}
\Volume{XX}
\Issue{1}
\Month{01}
\Year{2003}
\pagespan{3}{}
\Receiveddate{*}
\Reviseddate{**}
\Accepteddate{***}
\Dateposted{****} 



\title[The Be(0001) acoustic surface plasmon dispersion]
{Band structure effects on the Be(0001) acoustic-surface-plasmon
energy dispersion}


\author[V. M. Silkin et al.]{V. M. Silkin\footnote{Corresponding
     author: e-mail: {\sf waxslavs@sc.ehu.es}, Phone: +34\,943\,018\,284,
     Fax: +34\,943\,015\,600}\inst{1}} \address[\inst{1}]{Donostia International Physics Center (DIPC),
     Paseo de Manuel Lardizabal 4, 20018 San Sebasti\'an, Basque Country, Spain}
\author[]{J. M. Pitarke\inst{2,3}}
\address[\inst{2}]{CIC nanoGUNE Consolider, Mikeletegi Pasealekua 56, E-20009 Donostia, Basque Country, Spain}
\address[\inst{3}]{Materia Kondentsatuaren Fisika Saila, Zientzi Fakultatea, Euskal
Herriko Unibertsitatea, 644 Posta kutxatila, E-48080 Bilbo, Basque
Country, Spain}
\author[]{E. V. Chulkov\inst{1,4}}
\address[\inst{4}]{Departamento de F\'{\i}sica de Materiales and
Centro Mixto CSIC-UPV/EHU, Facultad de Ciencias Qu\'{\i}micas,
Universidad del Pa\'{\i}s Vasco UPV/EHU, Apto. 1072, 20080 San
Sebasti\'an, Basque Country, Spain} 
\author[]{B. Diaconescu\inst{5}}\address[\inst{5}]{Departament of Physics and Material Science Program,
University of New Hampshire, Durham, New Hampshire 03824, USA} 
\author[]{K. Pohl\inst{5}}
\author[]{L. Vattuone\inst{6}}\address[\inst{6}]{SNISM and Departimento di
Fisica, Universit\'a di Genova, 16146 Genova, Italy} 
\author[]{L. Savio\inst{6}}
\author[]{Ph. Hofmann\inst{7}}\address[\inst{7}]{Institute for Storage Ring Facilities and
Interdisciplinary Nanoscience Center (iNANO), University of Aarhus,
8000 Aarhus, Denmark}
\author[]{D. Far\'{\i}as\inst{8}}\address[\inst{8}]{Departamento de F\'{\i}sica
de la Materia Condensada, Universidad Aut\'onoma de Madrid, 28049
Madrid, Spain}
\author[]{M. Rocca\inst{9}}\address[\inst{9}]{IMEM-CNR and Departimento di
Fisica, Universit\'a di Genova, 16146 Genova, Italy}
\author[]{P. M. Echenique\inst{1,4}}
\begin{abstract}
We report first-principles calculations of acoustic surface plasmons on the (0001) surface of Be,
as obtained in the random-phase approximation of many-body theory. The energy dispersion of these
collective excitations has been obtained along two symmetry directions. Our results show a
considerable anisotropy of acoustic surface plasmons, and underline the capability of
experimental measurements of these plasmons to {\it map} the electron-hole excitation
spectrum of the quasi two-dimensional Shockley surface state band that is present on the Be(0001) surface.
\end{abstract}
\maketitle                   






\section{Introduction}

The existence of metal surfaces is well known to yield the appearance of conventional surface collective electronic excitations, surface plasmons, predicted
by Ritchie 50 years ago \cite{ripr57}. Thereafter, surface plasmons
were widely investigated both experimentally and theoretically~
\cite{fepss82,ra88,rossr95,pltsnimb95,li97,pisirpp07}. Another kind
of collective excitations that are present at metal surfaces are the so-called multipole surface plasmons predicted theoretically~\cite{beprb70} and later
detected experimentally~\cite{leplprl79,scscprb84}. The energy of both
these plasmons is a fraction of the bulk plasmon energy, $\omega_p$, typically
$(0.7-0.8)\omega_p$ for small momenta.

A qualitatively different surface collective excitation, called acoustic surface plasmon, has recently been predicted theoretically~\cite{sigaepl04} and observed
experimentally using angle-resolved electron energy loss spectroscopy~\cite{dipon07}. Acoustic surface plasmons are low-energy collective excitations that are confined to solid surfaces where a partially occupied quasi-two dimensional (2D)
surface-state band coexists with the underlying three-dimensional
(3D) continuum, as occurs in the case of Be(0001)~\cite{sigaepl04}
and the (111) surfaces of the noble metals Cu, Ag, and Au~\cite{sipiprb05}. They are called acoustic because their energy dispersion exhibits sound-like linear behaviour at low momenta, contrary to the square-root dispersion expected for a pure 2D electron gas~\cite{stprl67}.

The calculations that led to the prediction of acoustic surface
plasmons~\cite{sigaepl04} assumed translational invariance in the plane of the
surface and incorporated the electron dynamics in the direction
perpendicular to the surface through the use of a one-dimensional
(1D) model potential that describes the main features of the surface
band structure \cite{chsiss99}. Although acoustic plasmons had
previously been only expected to exist for spatially separated
plasmas~\cite{chjetp72,damaprb81}, 1D model potential
self-consistent calculations demonstrated that, contrary to
expectations, acoustic plasmons should indeed be present at bare
metal surfaces where a 2D surface-state band coexists with the
underlying 3D continuum.

In this paper, we report {\it first-principles} calculations of acoustic surface plasmons on the (0001) surface of Be, as obtained in the random-phase approximation of many-body theory. First-principles calculations along the $\Gamma K$ symmetry direction were reported before~\cite{dipon07}, but here we extend those calculations to other symmetry directions and focus on the discussion
of the anisotropy of acoustic surface plasmons. Our calculations show that the anisotropy is considerable and underline the capability of experimental measurements of these plasmons to {\it map} the electron-hole excitation spectrum of the quasi two-dimensional Shockley surface state band that is present on the Be(0001) surface.

\section{Calculation details}

The rate at which a frequency-dependent external potential
$\phi^{ext}({\bf r},\omega)$ generates electronic excitations in a
many-electron system can be obtained, within lowest-order
perturbation theory, as follows~\cite{li97}
\begin{equation}\label{g_definition}
w(\omega)=-2\,{\rm Im}\int{\rm d}{\bf r}\, n_{\rm ind}({\bf
r},\omega)\,\phi^{ext}({\bf r},\omega),
\end{equation}
where $n_{\rm ind}({\bf r},\omega)$ represents the induced electron density. In the case of a
periodic solid surface, one can write~\cite{pisirpp07}
\begin{equation}
w(\omega)=\sum_{\overline{\bf q}}^{\rm SBZ}w(\overline{\bf
q},\omega),
\end{equation}
where the sum extends over the surface Brillouin zone (SBZ) and
$w(\overline{\bf q},\omega)$ denotes the rate at which the external
potential generates electronic excitations of frequency $\omega$ and
parallel wave vector $\overline{\bf q}$:
\begin{equation}\label{w2}
w(\overline{\bf q},\omega)=-{2\over A}\,{\rm Im}\int{\rm d}z\,
n_{\rm ind}(z,\overline{\bf q},\omega)\,\phi^{ext}(z,\overline{\bf
q},\omega),
\end{equation}
$n_{\rm ind}(z,\overline{\bf q},\omega)$ and
$\phi^{ext}(z,\overline{\bf q},\omega)$ representing 2D Fourier
transforms of the induced electron density $n_{\rm ind}({\bf
r},\omega)$ and the external potential $\phi^{ext}({\bf r},\omega)$,
respectively, and $A$ being the normalization area. If the 2D
Fourier transform of the external potential is of the form
\begin{equation}\label{external}
\phi^{ext}(z,\overline{\bf q},\omega)={-2\pi\over
\overline{q}}\,{\rm e}^{\overline{q}z},
\end{equation}
the rate $w(\overline{\bf q},\omega)$ can be expressed in the
following form:
\begin{equation}\label{absorption}
w(\overline{\bf q},\omega)={4\pi\over \overline{q}\,A}{\rm
Im}\,g(\overline{\bf q},\omega),
\end{equation}
where $g(\overline{\bf q},\omega)$ represents the so-called surface
response function \cite{pezaprb85}
\begin{equation}\label{g_final}
g(\overline{\bf q},\omega)=-\frac{2\pi}{\overline{q}} \int {\rm
d}z\,{\rm d}z'\,\chi_{\overline{\bf G}=0\overline{\bf
G}'=0}(z,z',\overline{\bf q},\omega),
\end{equation}
$\chi_{\overline{\bf G}=0\overline{\bf G}'=0}(z,z',\overline{\bf
q},\omega)$ being the Fourier components of the density response
function of the interacting many-electron system. In the
random-phase approximation of many-body theory, these Fourier
coefficients are easily obtained from the Fourier
coefficients $\chi_{\overline{\bf G}\,\overline{\bf G}'}^{\rm
o}(z,z',\overline{\bf q},\omega)$ of the noninteracting density
response function by solving the following matrix equation:
\begin{equation}\label{chi_reciprocal}
\chi_{{\bf G}{\bf G}'}(\overline{\bf q},\omega)=\chi^{\rm o}_{{\bf
G}{\bf G}'}(\overline{\bf q},\omega) + \sum_{{\bf G}_1} \chi^{\rm
o}_{{\bf G}{\bf G}_1}(\overline{\bf q},\omega) \,\upsilon_{{\bf
G}_1}(\overline{\bf q})\, \chi_{{\bf G}_1{\bf G}'}(\overline{\bf
q},\omega),
\end{equation}
where $v_{\bf G}(\overline{\bf q})={4\pi\over |\overline{\bf q}+{\bf
G}|^2}$ represent the Fourier coefficients of the bare Coulomb
electron-electron interaction.

For positive frequencies, the imaginary part of the Fourier
coefficients $\chi_{{\bf G}{\bf G}'}^{\rm o}(\overline{\bf
q},\omega)$ of the noninteracting density response function is
easily obtained from the spectral function $S^{\rm o}_{{\bf G}{\bf
G}'}(\overline{\bf q},\omega)$, as follows
\begin{equation}\label{SGG}
{\rm Im}\left[\chi_{{\bf G}{\bf G}'}^{\rm o}(\overline{\bf
q},\omega)\right]=-\pi S^{\rm o}_{{\bf G}{\bf G}'}(\overline{\bf
q},\omega),
\end{equation}
where
\begin{eqnarray}\label{spectral_function}
S^{\rm o}_{{\bf G}{\bf G}'}(\overline{\bf q},\omega)&=& \frac{2}{A}
\sum^{\rm SBZ}_{\overline{\bf k}} \sum_{n}^{\rm occ} \sum_{n'}^{\rm
unocc} \langle\psi_{\overline{\bf k}n}|e^{-{\rm i}(\overline{\bf
q}\,\overline{\bf r}+{\bf G}{\bf r})}|\psi_{\overline{\bf
k}+\overline{\bf q}n'}\rangle \cr\cr\cr
&\times&\langle\psi_{\overline{\bf k}+\overline{\bf q}n'}|e^{{\rm
i}(\overline{\bf q}\,\overline{\bf r}+{\bf G}'{\bf
r})}|\psi_{\overline{\bf k}n}\rangle
\delta(\varepsilon_{\overline{\bf k}n}-\varepsilon_{\overline{\bf
k}+\overline{\bf q}n'}+\omega),
\end{eqnarray}
the factor 2 accounting for spin. The sum over $n$ ($n'$) runs
over occupied (unoccupied) states of the surface band structure for
2D wave vectors $\overline{\bf k}$ in the SBZ. We take $\varepsilon_{\overline{\bf k}n}$ and
$\psi_{\overline{\bf k}n}$ to be the eigenvalues and eigenfunctions
of a first-principles single-particle Kohn-Sham (KS) Hamiltonian of
density-functional theory (DFT). For the evaluation of the real part
of the Fourier coefficients $\chi_{{\bf G}{\bf G}'}^{\rm
o}(\overline{\bf q},\omega)$ of the noninteracting density response
function, we perform a Hilbert transform of the corresponding
imaginary part.

For a description of the (0001) face of hcp Be from first
principles, we employ a repeated-slabs geometry. The slabs
periodically repeat in the direction normal to the (0001) surface
and every slab consists from 24 atomic Be layers. The distance
between slabs is chosen to correspond to 8 Be atomic layers. For the
evaluation of the eigenvalues and eigenfunctions entering Eq.
(\ref{spectral_function}) we solve the Kohn-Sham equation of DFT
self-consistently with the use of norm-conserving
pseudopotentials~\cite{chsifmm87}, the local-density approximation
(LDA) for exchange and correlation, and the Perdew-Zunger
parametrization of the Ceperley-Alder diffusion Monte Carlo energies
of a uniform electron gas~\cite{cealprl80,pezuprb81}. We expand the
single-particle wave functions $\psi_{\overline{\bf k},n}({\bf r})$
in a plane wave basis, with reciprocal vectors ${\bf G}$ up to an
energy cut-off of 20 Ry. For the lattice constants, we take the
experimental values $a=2.285$ $\AA$ and $c=3.585$
$\AA$~\cite{amivfmm62}. The experimental value \cite{dahaprl92} for
the relaxation of surface atomic layers has been taken into account.
The sum over 2D wave vectors $\overline{\bf k}$ in
Eq.~(\ref{spectral_function}) is taken to run over a $108\times108$
mesh, which corresponds to 11664 points in the SBZ. The energy band
summation includes all bands up to an energy cut-off of 50 eV above
the Fermi level. The integrations involved in the evaluation of the
surface response function of Eq.~(\ref{g_final}) are performed from
$-c/2$ up to $c/2$, where $c$ is the unit cell size in the
$z$-direction. Finally, we note that we include the so-called
crystal local-field effects in the normal direction only and neglect
them along the surface, i.e., only the
$\overline{\bf G}=\overline{\bf G}'=0$
element of the noninteracting density response
matrix $\chi_{\overline{\bf G}\overline{\bf
G}'}^0(z,z',\overline{\bf q},\omega)$ has been considered in the evaluation of the
coefficients $\chi_{\overline{\bf G}=0\overline{\bf
G}'=0}(z,z',\overline{\bf q},\omega)$ of the interacting density
response function; lateral crystal local-field effects
were found to be negligible in the case of the Mg(0001) surface~\cite{sichprl04}, and we expect them to be negligibly small, as
well, in the case of other simple metal surfaces like Be(0001).

\section{Results and discussion}

\begin{figure}
\includegraphics[width=0.45\linewidth,angle=270]{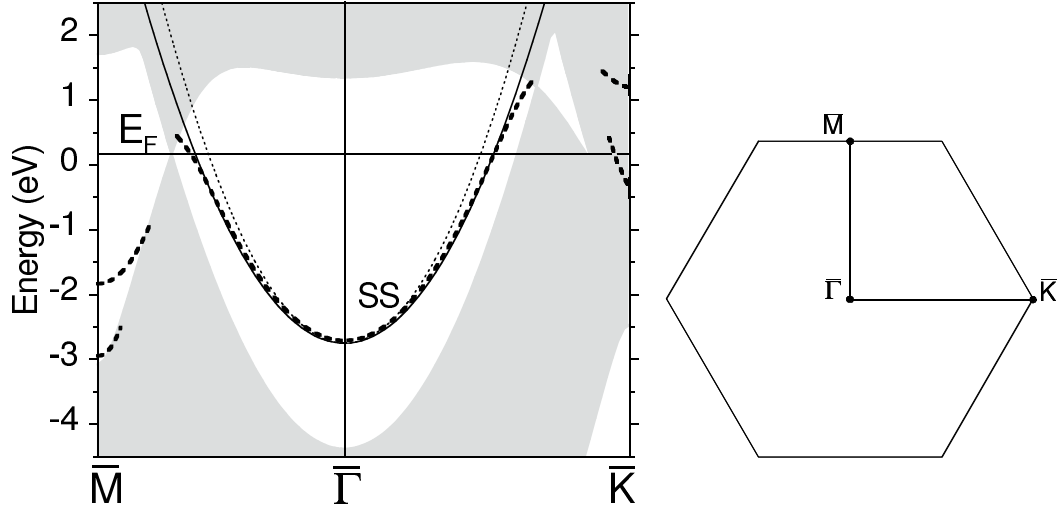}\\
\caption{ {\it Ab initio} band structure of the Be(0001) surface.
The Fermi level, ${\rm E}_{\rm F}$, is at the zero energy. The grey
areas show projected bulk Be electronic states. The surface states
dispersions are shown by dashed lines. Thin solid line shows
parabolic fit to the surface state SS located at the
$\overline{\Gamma}$ point with effective mass $m^*_{\rm SS}=1.2$.
The bare surface state dispersion with $m^*_{\rm SS}=1$ is shown by
thin dotted line. The surface Brillouin zone with two symmetry
directions, $\overline{\Gamma}\,\overline{\rm M}$ and
$\overline{\Gamma}\,\overline{\rm K}$, is shown on the right.}
\label{Be0001_band_structure}
\end{figure}

Figure~\ref{Be0001_band_structure} represents the projected
electronic structure of the Be(0001) surface, as obtained along the
symmetry directions $\overline{\Gamma}\,\overline{\rm K}$ and
$\overline{\Gamma}\,\overline{\rm M}$. The characteristic feature of
this surface is the presence of wide energy gaps in the projected
bulk band structure, where several strongly localized surface states
reside \cite{kaflssc84,bajeprb85,chsiss87}. Here we focus our
attention on the Shockley surface state that is located around the
SBZ center and is marked by SS. According to our present
first-principles calculations, the binding energy of this surface
state at $\overline{\Gamma}$ is 2.75 eV in close agreement with the
experimental
data~\cite{kaflssc84,bajeprb85,hefrprb99,lajeprb00,vofunimb06} and
other {\it ab initio} calculations~\cite{chsiss87,feprb92}. As for
the energy dispersion of this surface-state band, we find that
within the occupied part of the band it is almost isotropic and well
described by a parabolic-like dispersion with an effective mass
$m^*_{\rm SS}=1.2$, again in good agreement with
experiment~\cite{bajeprb85,pespprb98}. However, within the
unoccupied part, where the surface-state energy approaches the
borders of the energy gap, our calculated energy dispersion deviates
substantially from the quasi-free electron like parabolic behaviour.

\begin{figure}
\includegraphics[width=0.7\linewidth,angle=0]{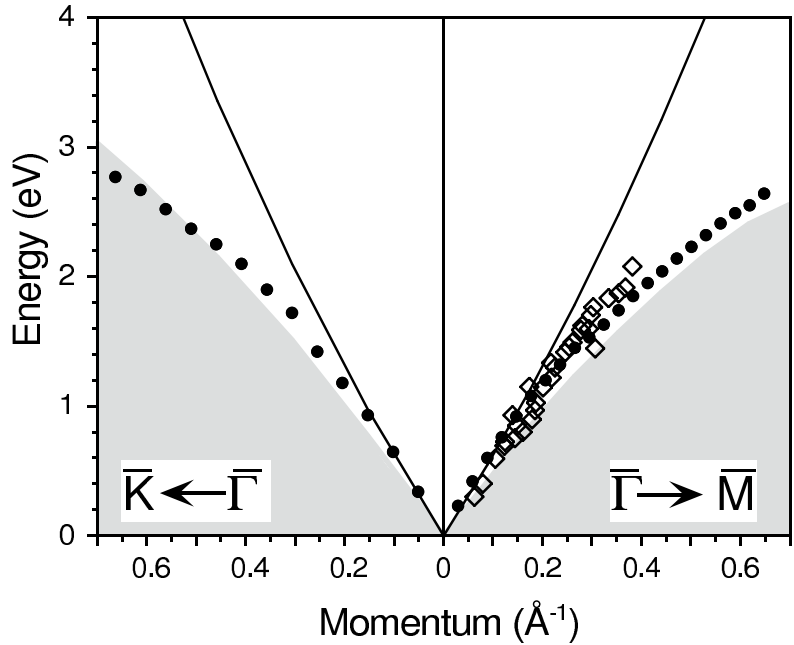}\\
\caption{ Calculated acoustic surface plasmon dispersion at the
Be(0001) surface along two symmetry directions,
$\overline{\Gamma}\,\overline{\rm M}$ and
$\overline{\Gamma}\,\overline{\rm K}$, is shown by dots. Grey areas
show the "momentum-energy" phase space where intraband electron-hole
excitations within the SS surface state are available. The
experimental data \cite{dipon07} along the
$\overline{\Gamma}\,\overline{M}$ direction are shown by open
diamonds. Thin solid line shows acoustic surface plasmon calculated
with the use of the Be(0001) model potential \cite{chsiss99} and SS
effective mass $m^*_{\rm SS}=1.2$.} \label{ASP_dispersion}
\end{figure}

Collective excitations created by an external potential of the form
dictated by Eq.~(\ref{external}) can be traced to the peaks of the
absorption probability $w(\overline{\bf q},\omega)$ of
Eq.~(\ref{absorption}), i.e., to the peaks of the imaginary part of
the surface response function $g(\overline{\bf q},\omega)$ of
Eq.~(\ref{g_final}). We have searched for the maxima of
$g(\overline{\bf q},\omega)$, as obtained as a function of the
frequency $\omega$, for various 2D $\overline{\bf q}$ wave vectors,
and we have found the low-energy (acoustic) surface-plasmon energy
dispersion plotted in Fig.~\ref{ASP_dispersion} along two distinct
symmetry directions: $\overline{\Gamma}\,\overline{\rm M}$ and
$\overline{\Gamma}\,\overline{\rm K}$.

As already shown by the 1D model potential calculations reported
before \cite{sigaepl04}, the energy dispersion of acoustic surface
plasmons follows very closely the upper edge of the continuum
(represented in Fig.~\ref{ASP_dispersion} by the grey area) of 2D
electron-hole pair excitations occurring within the quasi-2D
surface-state band. This continuum of electron-hole pair excitations
(and, in particular, its upper edge) is, however, dictated by the
energy dispersion of the surface-state band, which strongly depends
on the actual band structure and deviates substantially from a
quasi-free electron-like parabolic behaviour, as discussed above.
This results in a first-principles energy dispersion of acoustic
surface plasmons that deviates from previous 1D model potential
calculations, shows excellent agreement with the available
experimental data along the $\overline{\Gamma}\,\overline{\rm M}$
direction, and exhibits notable anisotropy.

We attribute our predicted anisotropy of acoustic surface plasmons to
the fact that the upper edge of the 2D electron-hole pair
excitations occurring within the quasi-2D surface-state band depends
strongly on the symmetry direction. Our calculations also show that
the peaks in the surface loss function are stronger and narrower
along the $\overline{\Gamma}\,\overline{\rm M}$ than in the case of
the $\overline{\Gamma}\,\overline{\rm K}$ direction. We attribute
this fact to the complicate Be(0001) surface electronic structure,
where additionally inter-band transitions involving other surface
states are also possible (see Fig. \ref{Be0001_band_structure}). It
seems that the destructive role of such transitions on the
properties is more pronounced in the
$\overline{\Gamma}\,\overline{\rm K}$ direction. Clearly there is no
anisotropy in the acoustic surface plasmon dispersion obtained with the use of a 1D model potential~\cite{sigaepl04}, simply due to the isotropic surface
electronic structure inherent to that approximation. For small momenta, our {\it ab initio} acoustic surface plasmon dispersion almost coincides with the
dispersion obtained with the use of the 1D model potential and $m^*_{\rm SS}=1.2$. As the momentum increases, the two calculated curves start to
deviate as a consequence of the deviation in the corresponding surface
state dispersions obtained in the two models. This demonstrates the importance of the unoccupied part of the surface state dispersion, as it determines the upper edge for the intra-band electron-hole excitation continuum, which represents the crucial factor in the determination of the acoustic-surface-plasmon energy dispersion~\cite{pinaprb04}.

\section{Summary and conclusions}

In summary, we have reported first-principles calculations of
acoustic surface plasmons on the (0001) surface of Be, as obtained
in the random-phase approximation of many-body theory. We have
calculated the energy dispersion of these collective excitations
along the symmetry directions $\overline{\Gamma}\,\overline{\rm M}$
and $\overline{\Gamma}\,\overline{\rm K}$, and we have found that the energy
dispersion (i) follows closely the upper edge of the continuum of 2D
electron-hole pair excitations occurring within the quasi-2D
Shockley surface state band that is present in this surface, (ii)
deviates considerably from previous 1D model potential calculations,
(iii) shows excellent agreement with the available experimental data
along the $\overline{\Gamma}\,\overline{\rm M}$ direction, and (iv)
exhibits notable anisotropy. Because the energy dispersion of
acoustic surface plasmons follows closely the upper edge of the
continumm of 2D electron-hole pair excitations occurring within the
quasi-2D Shockley surface state band, we conclude that experimental
measurements of these plasmons serve to {\it map} the anisotropic
electron-hole excitation spectrum of the surface-state band that is
present on the Be(0001) surface. The same conclusion should be
applicable to other surfaces containing a partially Shockley surface
state band near the Fermi level, as occurs in the case of the (111)
surfaces of the noble metals Cu, Ag, and Au.

\begin{acknowledgement}
V.M.S., J.M.P, E.V.C, and P.M.E. acknowledge partial support
from the University of the Basque Country (9/UPV
00206.215-13639/2001), the Basque Unibertsitate eta Ikerketa Saila, and the Spanish Ministerio de Educaci\'on y Ciencia (MEC)
(FIS 2004-06490-C03-01 and CSD2006-53). B.D. and K.P. acknowledge the
National Science Foundation; L.V., L.S., and M.R. Compagnia di San
Paolo; P.H. the Danish Natural Science Research Council; D.F. the
Programa Ramon y Cajal and Comunidad de Madrid.
\end{acknowledgement}


\end{document}